\documentclass[reprint]{JASA}

\usepackage{booktabs}
\usepackage{multirow}
\usepackage{textcomp}
\usepackage{siunitx}
\hypersetup{hypertexnames=false}

\newcommand{\ci}[1]{\ensuremath{^{\scriptscriptstyle \pm\,#1}}}

\begin{document}


\title[Navigating Speech Enhancement for Real-Time MRI]{Navigating Speech Enhancement for Real-Time MRI: A Systematic Assessment of Signal Quality, Source Preservation, and Downstream Tasks}

\author{Huang-Cheng Chou}
\affiliation{Ming Hsieh Department of Electrical and Computer Engineering, University of Southern California, Los Angeles, California 90089, United States}

\author{Sean Foley}
\affiliation{Ming Hsieh Department of Electrical and Computer Engineering, University of Southern California, Los Angeles, California 90089, United States}
\affiliation{Department of Linguistics, University of Southern California, Los Angeles, California 90089-1693, United States}

\author{Haley Hsu}
\affiliation{Department of Linguistics, University of Southern California, Los Angeles, California 90089-1693, United States}

\author{Kevin Huang}
\affiliation{Ming Hsieh Department of Electrical and Computer Engineering, University of Southern California, Los Angeles, California 90089, United States}

\author{Szu-Jui Chen}
\affiliation{Erik Jonsson School of Engineering and Computer Science, University of Texas at Dallas, Richardson, Texas 75080-3021, United States}

\author{Rong Chao}
\affiliation{Research Center for Information Technology Innovation, Academia Sinica, Taipei 115201, Taiwan}

\author{Louis Goldstein}
\affiliation{Department of Linguistics, University of Southern California, Los Angeles, California 90089-1693, United States}

\author{Khalil Iskarous}
\affiliation{Department of Linguistics, University of Southern California, Los Angeles, California 90089-1693, United States}

\author{Dani Byrd}
\affiliation{Department of Linguistics, University of Southern California, Los Angeles, California 90089-1693, United States}

\author{Yu Tsao}
\affiliation{Research Center for Information Technology Innovation, Academia Sinica, Taipei 115201, Taiwan}

\author{Sudarsana Reddy Kadiri}
\affiliation{Ming Hsieh Department of Electrical and Computer Engineering, University of Southern California, Los Angeles, California 90089, United States}

\author{John H. L. Hansen}
\affiliation{Erik Jonsson School of Engineering and Computer Science, University of Texas at Dallas, Richardson, Texas 75080-3021, United States}

\author{Shrikanth Narayanan}
\email{shri@usc.edu}
\affiliation{Ming Hsieh Department of Electrical and Computer Engineering, University of Southern California, Los Angeles, California 90089, United States}

\begin{abstract}
Audio recorded during real-time magnetic resonance imaging (rtMRI) is heavily contaminated by scanner noise, but it remains unclear whether general-purpose speech enhancement improves the signal for speech research and downstream processing. Three off-the-shelf systems---Denoiser, PASE, and RE-USE---are evaluated across five rtMRI corpora using naturally recorded inputs, a clean-input probe, and an archived paired additive-noise probe. The multi-task evaluation spans learned quality predictors, speaker and phone representations, reference-based intelligibility and quality measures, acoustic--phonetic probes, automatic speech recognition (ASR), and paralinguistic tasks. The central result is that enhancement effects are endpoint dependent: higher predicted-quality scores do not reliably imply better ASR performance or greater source fidelity. Across 15 corpus--recognizer comparisons using corpus-provided processed inputs, RE-USE yielded lower word-error-rate point estimates in 11, whereas Denoiser yielded higher estimates in 13. In the paired additive-noise probe, PASE and RE-USE improved recognized-phone agreement, intelligibility, and perceptual-quality point estimates. Denoiser improved recognized-phone agreement and short-time objective intelligibility (STOI) but reduced speaker-embedding similarity. No system was uniformly best across corpora, recognizers, and endpoints. Enhanced rtMRI audio should therefore be treated as a task-specific transformed derivative rather than a universally improved replacement for the original or DSP-processed waveform.
\end{abstract}

\keywords{Real-Time Magnetic Resonance Imaging (rtMRI); speech enhancement; acoustic phonetics; downstream task utility; speech corpus;multi-corpus benchmark}

\maketitle

\section{Introduction}
Real-time magnetic resonance imaging (rtMRI) has become a powerful modality for observing the moving vocal tract during running speech, capturing the coordinated motions of the tongue, lips, velum, and pharynx that are difficult to access with other instruments \cite{Narayanan_2014, Kumar_2024}.
Over the past decade, several public rtMRI corpora have been released, e.g., \cite{Narayanan_2014, Kim_2014, Lim_2021, Foley_2026}, providing rich resources for research on multi-lingual speech production, language development, and speech disorders.
A key feature of these rtMRI corpora is that the speech audio is recorded simultaneously with the imaging, pairing articulatory dynamics during speech with the corresponding acoustic signal.
However, this co-recorded audio is severely contaminated by the intense, periodic acoustic noise produced by the MRI scanner, in addition to other noise sources in the scanner \cite{Bresch_2006}, which limits its broad application in spoken language research and technology development.

Specifically, residual scanner noise constrains reuse of the acoustic channel, even after preprocessing intended to suppress it.
For example, the USC Long-form Single-Speaker (LSS) \cite{Foley_2026} \textbf{Raw} recordings receive substantially lower scores than an unpaired clean \textbf{Lab} comparator from four learned, reference-free quality predictors, as quantified in the analyses below.
Such degradation can affect downstream tasks including automatic speech recognition (ASR), speaker modeling, speech emotion recognition (SER), and age estimation.

Recent advances in speech enhancement offer an opportunity to revisit these rtMRI corpora and improve the usability of their acoustic channels.
Modern speech enhancement models can suppress strong, non-stationary noise while aiming to preserve linguistic and paralinguistic content \cite{Defossez_2020, Rong_2026, Fu_2026}.
Applying such models to rtMRI audio could, in principle, open these corpora to a broader range of end-use applications.
However, it remains unclear whether enhanced rtMRI speech improves downstream performance or whether enhancement-induced changes alter linguistic and paralinguistic cues that are important for analysis.
Answering this question requires a systematic, multi-task evaluation rather than reliance on learned speech-quality predictors alone.
Because speech is a complex, multi-faceted signal carrying acoustic, phonetic, speaker, and paralinguistic information, no single metric can capture overall processing effects for all research applications. Evaluating speech enhancement therefore requires complementary tasks and measures that span signal-level quality, acoustic-phonetic probes, speaker representations, and downstream machine tasks.

We organize the current study around four questions. 
\begin{itemize}
    \item (RQ1) examines how off-the-shelf enhancement changes reference-free predicted quality and speaker-related, recognized-phone, modulation, and formant measures. 
\item (RQ2) evaluates how enhancement affects ASR across corpora and recognizers. 
\item (RQ3) investigates how enhancement affects paralinguistic classification, specifically SER and sex/gender-related classification probes. 
\item (RQ4) explores whether enhancement effects differ by age group and first-language group and whether age-related information is retained.
\end{itemize}
 
The contribution is not a new enhancement algorithm but rather a multi-corpus evaluation framework that connects predicted signal quality with acoustic, phonetic, speaker-related, and downstream measures under naturally recorded, clean-input, and archived paired conditions.
Our main contributions in this work are as follows:
\begin{itemize}
    \item We evaluate three enhancement systems on four public and one internal rtMRI corpus, comparing Raw audio, the corpus-provided digital signal processing (DSP) baseline described in Section~\ref{ss:input_conditions}, clean Lab audio, and model-enhanced audio where each condition is available.
    \item We find that higher reference-free predicted-quality scores do not consistently correspond to better ASR or speaker-embedding outcomes; the acoustic--phonetic probes reveal distinct, model-dependent changes in their target constructs.
    \item We use clean-input and archived additive-noise probes to separate processing-induced changes from performance under scanner-noise contamination.
    \item We report descriptive age- and first-language-group analyses that expose substantial corpus and model dependence, while reporting marginal confidence intervals and explicitly documenting sampling limitations.
\end{itemize}

\section{Related work}
\subsection{Real-time MRI speech corpora and their uses}
Public rtMRI speech corpora have been developed primarily to support research on speech production, with synchronized midsagittal image sequences of the vocal tract serving as the principal data modality and the accompanying audio typically serving a complementary role~\cite{Narayanan_2014,Kim_2014,Lim_2021}.
Accordingly, research using these corpora has largely addressed articulatory and image-based questions, including vocal-tract image segmentation~\cite{Bresch_2009,Hebbar_2020}, vocal-tract constriction dynamics, inter- and intra-speaker variability in articulation~\cite{Ramanarayanan_2018,Sorensen_2019,Lammert_2013}, the articulation of emotional speech~\cite{Kim_2014}, and more recently, machine-learning approaches to image reconstruction, for which the limited availability of large public datasets remains a recognized constraint~\cite{Lim_2021,Kumar_2024}.
In such studies, the concurrently recorded audio has commonly supported transcription, forced alignment, and the temporal interpretation of articulatory events rather than serving as the primary signal of interest.

However, despite the use of custom fiber-optic microphones and specialized adaptive noise-cancellation pipelines, these recordings remain contaminated by MRI gradient noise~\cite{Bresch_2006,Vaz_2018,Narayanan_2014,Lim_2021}.
MRI gradient noise is intense and strongly time structured, typically comprising repeated broadband transients and harmonic components synchronized with the imaging sequence. Its spectrum and level depend on the pulse sequence and gradient operation as well as the scanner and recording configuration. Among the documented acquisitions in this study, the Long-form Single-Speaker corpus used a $0.55$-T scanner, whereas the three other public corpora described below used $1.5$-T scanners. Because field strength is confounded with acquisition and preprocessing differences, we do not attribute multi-corpus acoustic differences to field strength alone.

Prior rtMRI work has emphasized synchronized acquisition and corpus-specific scanner-noise reduction, including noise-robust recording and adaptive cancellation~\cite{Bresch_2006} and spectral--temporal dictionary learning~\cite{Vaz_2018}. General enhancement benchmarks, by contrast, primarily evaluate perceptual quality and intelligibility on conventional noisy speech. Neither line of work establishes whether an off-the-shelf enhancement front end preserves measurement-relevant signal properties across rtMRI corpora with different scanners, microphones, and released preprocessing. We address this gap by combining multi-corpus downstream evaluation with clean-input and archived paired additive-noise probes that distinguish processing-induced changes from performance under scanner-noise contamination. Exploratory analyses additionally examine children's speech and speakers grouped by first-language metadata.

\subsection{Speech enhancement and downstream evaluation}
Modern speech enhancement spans predictive/causal denoisers~\cite{Defossez_2020}, generative approaches that resynthesize clean speech~\cite{Rong_2026}, and universal/foundation enhancement models trained across diverse conditions~\cite{Fu_2026}.
Although these models can increase human-perceived quality or scores from learned quality predictors, such gains do not necessarily translate into better downstream machine performance: ASR may show no corresponding improvement and can even deteriorate when enhancement alters recognition-relevant acoustic cues. An unimproved recognition result is not intrinsically harmful, but it shows that the enhancement front end supplied no benefit for that task despite its computational cost and potential signal modification. This motivates evaluation beyond signal-level quality.
This concern is sharpened by the increasing use of non-intrusive, learned predictors of mean opinion score (MOS), such as UTMOS~\cite{Saeki_2022}, UTMOSv2~\cite{Baba_2024}, VQScore~\cite{Fu_2024}, and SHEET SSL-MOS~\cite{Huang_2025}, whose validity for out-of-domain material such as noisy rtMRI and child speech has not been established.
We therefore evaluate enhancement with a multi-task battery comprising learned quality predictors, speaker-embedding and recognized-phone measures, acoustic-modulation and formant-based probes, ASR, emotion recognition, sex/gender-related classification, and speaker age estimation.

\section{Speech corpora}
We evaluate speech enhancement on five rtMRI speech corpora. 
Four of them are publicly available: USC Long-form Single-Speaker (\textbf{LSS}) \cite{Foley_2026}, USC-TIMIT (\textbf{TIMIT}) \cite{Narayanan_2014}, USC 75-Speaker (\textbf{75-Spk}) \cite{Lim_2021}, and USC-EMO-MRI (\textbf{EMO-MRI}) \cite{Kim_2014}. The fifth, the Child Speech and Adult Controls corpus (\textbf{Child}), was collected at USC and is available through an application-based access process.
For brevity we refer to the corpora by these abbreviations throughout. 
Together they span a wide range of speakers, ages, speaking styles, and recording configurations. 
Table~\ref{tab:datasets} summarizes the statistics of the audio actually processed in this study.

\begin{table*}[!t]
  \centering
  \caption{Five rtMRI corpora. ``\#Spk (F:M)'': speakers (female:male metadata); ``Dur.'': hours (mean file length, s); ``Input SR'': study-input rate. File counts may include multiple items. Tables II--XI use \textbf{bold} for best and \underline{underlining} for second-best values; displayed ties share rank, and arrows show direction.}
  \label{tab:datasets}
  \footnotesize
  \setlength{\tabcolsep}{3pt}
  \begin{tabular*}{\textwidth}{@{\extracolsep{\fill}}l c r r c c@{}}
    \toprule
    Corpus & \shortstack{\#Spk\\(F:M)} & \#Files & \shortstack{Dur.\ hrs\\(avg.\ s)} & Age (years) & \shortstack{Input SR\\(kHz)} \\
    \midrule
    LSS       & 1 (0:1)   & 71   & 0.93 (47.30--47.40)\textsuperscript{a}  & 32      & 16 \\
    TIMIT & 10 (5:5)  & 1284\textsuperscript{b} & 8.52 (23.90)  & 20--46      & 20 \\
    75-Spk\textsuperscript{c}     & 75 (40:35)& 2371 & 20.71 (31.40) & 18--59  & 20 \\
    EMO-MRI     & 10 (5:5)  & 2304 & 2.36 (3.70)   & --      & 20 \\ \midrule
    Child & \shortstack{11 (5:6) children\\3 (1:2) adult controls} & 450 & 5.06 (40.50) & \shortstack{6.75--8.67 (children)\\22.83--30.58 (adult controls)} & 16 \\
    \bottomrule
  \end{tabular*}
  \begin{minipage}{\textwidth}
    \footnotesize\raggedright
    \textsuperscript{a}For LSS, the mean file length is 47.30~s in the updated Raw branch and 47.40~s in the DSP branch.\par
    \textsuperscript{b}The TIMIT count combines 914 rtMRI-session files and 370 separate Lab-session files.\par
    \textsuperscript{c}The 75-Spk audio was acquired at $100$ kHz and low-pass filtered and decimated to the $20$-kHz input waveform.
  \end{minipage}
\end{table*}

\subsection{USC Long-form Single-Speaker (LSS)}
\label{sec: usc_lss}
The USC Long-form Single-Speaker (LSS) dataset \cite{Foley_2026} is a long-form rtMRI speech dataset from a single speaker. 
The full corpus contains 684 utterances totaling roughly $54$ minutes of speech from a single male speaker of American English, including $37$ minutes of read speech and $17$ minutes of spontaneous speech. 
Among these utterances is the complete set of sentences used in the TIMIT corpus (Section~\ref{sec: usc_timit}). 
Data were acquired on a $0.55$-T scanner with a custom upper-airway receiver coil. 
The acoustic signal was sampled at $16$ kHz and the rtMRI video was reconstructed at $99$ frames/sec \cite{Kumar_2024}; hand-corrected phoneme alignments are also provided. 
In this study we process the long-form recordings listed in Table~\ref{tab:datasets}.

\subsection{USC-TIMIT (TIMIT)}
\label{sec: usc_timit}
USC-TIMIT \cite{Narayanan_2014} is a multimodal database for speech production research with separate rtMRI and electromagnetic articulography (\textbf{EMA}) sessions, each accompanied by a synchronously recorded acoustic channel. 
We use speech from 10 native speakers of General American English ($5$ female, $5$ male), each reading the same $460$-sentence MOCHA-TIMIT set, which is designed to elicit all English phonemes across a wide range of phonological contexts.
The processed data comprise 914 files from the rtMRI acoustic channel and 370 from the separate Lab session.
The rtMRI data were acquired on a $1.5$-T GE Signa Excite HD scanner with a custom four-channel upper-airway receiver coil. 
A fiber-optic microphone recorded audio at $20$ kHz in synchronization with the scanner's $10$-MHz clock; the released acoustic signal from the rtMRI session underwent custom adaptive periodic-noise cancellation.

\subsection{USC 75-Speaker (75-Spk)}
The USC 75-Speaker corpus \cite{Lim_2021} is a multispeaker dataset of raw and reconstructed rtMRI video, 3D volumetric images, and simultaneously recorded speech. 
The source corpus description reports $75$ speakers ($40$ female, $35$ male; $49$ classified as native English speakers and $26$ as non-native English speakers) spanning ages $18$--$59$, making it the most demographically diverse corpus in our study. 
Speakers produced both scripted speech (e.g., vowel--consonant--vowel (VCV) sequences, /bVt/ words, \& reading passages) and spontaneous speech (picture descriptions \& open-ended topics).
The rtMRI data were acquired on a $1.5$-T GE Signa Excite scanner with a custom eight-channel upper-airway receiver coil and a fiber-optic microphone. 
Audio was acquired at $100$ kHz in synchronization with the scanner's $10$-MHz clock, then low-pass filtered, decimated to $20$ kHz, and processed with normalized least-mean-square noise cancellation. 
This corpus contributes the largest amount of audio among the five corpora.

\subsection{USC-EMO-MRI (EMO-MRI)}
USC-EMO-MRI \cite{Kim_2014} is an emotional speech production database recorded with rtMRI. 
It contains recordings of $10$ professional actors ($5$ female, $5$ male), each producing the ``Grandfather'' passage and a set of seven sentences under four acted emotions: neutrality, anger, happiness, and sadness. 
Per-utterance perceptual emotion labels from at least ten listeners are provided. 
The recordings were acquired on a $1.5$-T scanner with a custom upper-airway receiver coil and a custom fiber-optic microphone sampled at $20$ kHz in synchronization with imaging; the released audio underwent adaptive noise cancellation. This corpus has the shortest mean utterance length in our study.

\subsection{Child speech and adult controls (Child)}
We also use an unreleased rtMRI speech corpus, hereafter referred to as \textbf{Child}, comprising two cohorts.
The child cohort includes 11 children (5 female, 6 male; mean age = $7.78$ years, standard deviation (SD) = $0.73$, range = $6.75$--$8.67$ years).
The adult control cohort includes 3 adults (1 female, 2 male; mean age = $25.75$ years, SD = $4.22$, range = $22.83$--$30.58$ years).
Throughout the analyses, \emph{Overall} denotes the two cohorts combined, whereas \emph{Children} and \emph{Adult Controls} denote the age-stratified subsets.
The corpus was collected using the same acquisition protocol and scanner as the LSS corpus (Section~\ref{sec: usc_lss}).
Participants produced sentences targeting a range of phonetic material (e.g., fricatives \& liquids) in a repeat-after-me paradigm, as well as spontaneous speech elicited using picture-based prompts.
The audio was originally sampled at $16$ kHz.

\section{Speech enhancement models and input conditions}

\subsection{Selection of speech enhancement models}
We evaluate three state-of-the-art, publicly available, pretrained speech enhancement models with complementary architectures for enhancing rtMRI audio.  We apply all models off-the-shelf, without any fine-tuning on rtMRI data, and operate on $16$ kHz mono audio, resampling inputs and outputs as needed. The three enhancement models are:
\begin{itemize}
    \item \textbf{Denoiser}~\cite{Defossez_2020}\footnote{\url{https://github.com/facebookresearch/denoiser}}: a causal waveform-domain encoder--decoder with a long short-term memory (LSTM) bottleneck, trained with time- and frequency-domain losses. We use the released causal Deep Noise Suppression (DNS) Challenge-trained \texttt{dns64}~\footnote{The checkpoint file is \path{dns64-a7761ff99a7d5bb6.th}.}, whose initial hidden-channel width is $H=64$; it operates on $16$-kHz mono audio.
    \item \textbf{PASE}~\cite{Rong_2026}\footnote{\url{https://github.com/cisco-open/pase}}: a generative system combining a WavLM-based denoising component with a dual-stream vocoder~\footnote{We use the released checkpoint \texttt{DeWavLM.tar} for denoising and \texttt{Vocoder\_Dual.tar} for synthesis. Both are from Hugging Face revision \texttt{efc471630e9e}.}.
    \item \textbf{RE-USE}~\cite{Fu_2026,NVIDIA_2026}: a bidirectional Mamba-based enhancement model trained on multiple degradation types under a distortion--perception objective~\footnote{We use the released model from Hugging Face revision \texttt{fe51d6495e49}. The model card states that this public artifact differs from the checkpoint evaluated in the accompanying preprint; our RE-USE results therefore apply to the released revision rather than constituting a reproduction of the preprint's checkpoint.}.
\end{itemize}
This selection contrasts a causal discriminative system with two recent generative or broadly trained systems. The final Denoiser condition uses the same \texttt{dns64} ($H=64$) checkpoint and inference procedure throughout. Each input waveform was divided into consecutive segments of at most 60~s. For every segment after the first, the preceding 2~s of input was prepended as left context and the corresponding context samples were discarded from the model output; the retained segments were then concatenated without cross-fading. Outputs were peak-limited only when their absolute peak exceeded one; files of at most 60~s were processed in a single segment. The LSS-DSP and Child Denoiser outputs and their downstream analyses were regenerated with this procedure; archived outputs from other long-file procedures were not used.
None of the enhancement models were fine-tuned in this study; therefore, the results characterize the specific pretrained checkpoints and inference pipelines evaluated here, rather than the model families in general.

\subsection{Input conditions}
\label{ss:input_conditions}
The acoustic signals evaluated in this study had undergone various preprocessing pipelines prior to enhancement, depending on whether they were obtained directly from public corpus releases or retrieved from local laboratory archives.
To distinguish those inputs from modern speech enhancement output, we use the following condition labels:
\begin{itemize}
    \item \textbf{Raw}: for the LSS dataset, the original rtMRI audio with no noise cancellation applied.
    \item \textbf{DSP}: the pre-enhancement rtMRI signal available after corpus-provided or locally archived processing. Implementations differ among the corpora and include adaptive noise cancellation where documented. For LSS, \emph{post-DSP} is a condition label for the time-aligned representation in our local archive; the available corpus documentation does not identify its algorithm or parameters, so it should not be interpreted as a specific published pipeline.
    \item \textbf{Lab}: clean acoustic speech recorded during a separate EMA session for USC-TIMIT speakers producing the same sentence prompts. Here, \emph{Lab} refers only to the acoustic channel recorded during the EMA session, not to the EMA sensor trajectories. Because the Lab and rtMRI recordings are separate productions, Lab is an unpaired clean comparator for rtMRI audio; it is an exact reference only when that same Lab waveform is processed directly or used to construct the archived additive-noise probe.
\end{itemize}

Where both Raw and DSP are available, we apply each speech enhancement model to the Raw and DSP signals to test which input condition benefits more from enhancement. The LSS corpus provides both Raw and DSP conditions, whereas the remaining corpora are enhanced from the DSP condition.

\section{Evaluation framework and metrics}

\subsection{Reference-free speech-quality prediction}
\label{ss:sq_metrics}
We first assess model-predicted speech quality with four reference-free metrics because rtMRI corpora generally lack paired clean reference recordings:
\begin{itemize}
    \item \textbf{UTMOS}~\cite{Saeki_2022}\footnote{\url{https://github.com/tarepan/SpeechMOS}}: an ensemble MOS predictor built on self-supervised speech representations that estimates naturalness on a $1$--$5$ scale\footnote{We use \path{tarepan/SpeechMOS:v1.2.0} with the released \path{utmos22_strong} model.}.
    \item \textbf{UTMOSv2}~\cite{Baba_2024}\footnote{\url{https://github.com/sarulab-speech/UTMOSv2}}: an improved MOS predictor that fuses self-supervised features with a spectrogram-based deep image classifier via transfer learning\footnote{We use version \texttt{1.3.1.dev0} with model \texttt{fusion\_stage3/fold0/seed42}.}. We use its raw regression outputs without clipping to the nominal MOS range.\footnote{Because the released inference pipeline samples a random crop, we evaluate one crop per waveform using a seed derived from its canonical path relative to the SPAN data root. Before each separate prediction call, we reset the Python, NumPy, PyTorch, and CUDA random-number generators; inference uses batch size 1 and no data-loader workers. We verified that silent-section removal was not applied by the evaluated release, so the scores reflect its untrimmed-waveform behavior. This protocol makes each path-specific single-crop score reproducible in the fixed evaluation environment and invariant to processing order; it is not a full-waveform average. Because the seed includes the condition-specific path, corresponding conditions need not sample the same temporal crop; the paired bootstrap pairs scores by source recording but does not make the sampled crops identical.}
    \item \textbf{VQScore}~\cite{Fu_2024}\footnote{\url{https://github.com/JasonSWFu/VQscore}}: a self-supervised quality estimate derived from the quantization error of a vector-quantized autoencoder trained only on clean speech ($0$--$1$).
    \item \textbf{SHEET SSL-MOS}~\cite{Huang_2025}\footnote{\url{https://github.com/unilight/sheet}}: a MOS predictor trained and benchmarked with the heterogeneous MOS-Bench collection. We use SHEET\footnote{We use SHEET v0.2.5 with the released WavLM-Large checkpoint \texttt{bvcc+somos+singmos+}\linebreak[4]\texttt{nisqa+tmhint-qi+}\allowbreak\texttt{tencent+pstn+}\allowbreak\texttt{urgent2024-mos/}\allowbreak\texttt{sslmos-wavlm\_large/1337} at model-repository revision \texttt{2a5bf30f}.}.
    Inputs are mono at 16~kHz and inference uses batch size 1. Waveforms up to 30~s are scored whole; longer files are split into contiguous, non-overlapping 30-s segments with full coverage, and segment scores are averaged with duration weights. 
\end{itemize}
For all four metrics, higher values indicate a higher score from the corresponding learned predictor. 
These scores are not substitutes for listener judgments. 
The results reported here use the four predictors for the LSS Raw-input branch and for the clean-input and archived additive-noise probes described below. 

\subsection{Acoustic and representation-based probes}

\subsubsection{Speaker-embedding and recognized-phone analyses}

A central concern is whether speech enhancement alters speaker- and content-related information. 
We probe this using four TIMIT speakers (F1, F5, M1, M3) with Lab audio recorded during a separate EMA session. 
These Lab waveforms supply the input and reference for the clean-input and synthetic additive-noise probes; although they contain the same sentence texts, they are not paired references for the speakers' recorded rtMRI productions. 
In addition to the non-intrusive metrics above, we follow the metric implementations in the URGENT Challenge toolkit\footnote{\url{https://github.com/urgent-challenge/urgent2025_challenge}} and use two reference-based metrics:
\begin{itemize}
    \item \textbf{SpkSim}: the cosine similarity between speaker embeddings extracted by a RawNet3 speaker-verification model~\cite{Jung_2022}\footnote{\url{https://huggingface.co/espnet/voxcelebs12_rawnet3}} (range $-1$ to $1$). It quantifies consistency in one pretrained embedding space; without an impostor-score distribution, verification threshold, or listener judgment, it is not a direct test of perceived speaker identity.
    \item \textbf{LPS} (Levenshtein Phoneme Similarity)~\cite{Pirklbauer_2023}: one minus the length-normalized Levenshtein distance between phone sequences recognized from the enhanced and reference signals by a wav2vec2 phone recognizer.\footnote{\url{https://huggingface.co/facebook/wav2vec2-lv-60-espeak-cv-ft}} LPS is a recognizer-derived proxy for phone-sequence agreement, not a complete measure of linguistic-content preservation.
\end{itemize}

\paragraph{Clean-input probe.}
We first apply each speech enhancement model directly to the clean Lab audio and evaluate 
(i) speaker-embedding and recognized-phone agreement (SpkSim and LPS) against the original Lab waveform and 
(ii) reference-free predicted quality (UTMOS, UTMOSv2, VQScore, and SHEET). 
This probes how the models transform speech that is already clean.

\paragraph{Archived additive-noise probe.}
The clean-input probe isolates changes to already clean speech, but it does not measure enhancement under scanner-noise contamination. Because recorded rtMRI speech audio has no paired clean waveform, we analyze an archived author-generated probe in which the clean Lab waveforms of the same four TIMIT speakers were paired with files labeled as synthetic additive MRI-noise mixtures. The mixture-generation code and configuration were not retained, so the noise-recording source, segment assignment, random seed, gain normalization, and intended target signal-to-noise ratio (SNR) cannot be reconstructed. We therefore characterize the stored mixtures directly. After truncating each noisy--clean pair to its common duration, we define full-waveform SNR as $10\log_{10}\!\left[\sum_t s_t^2/\sum_t(y_t-s_t)^2\right]$, where $s_t$ is the clean Lab signal and $y_t$ is the stored mixture. Across 370 pairs, the realized median SNR is $-14.54$ dB (range $-36.40$ to $-2.30$ dB). 
Across all stored mixtures, the maximum absolute normalized sample value was $0.728$, so none contained samples at digital full scale. Each archived mixture has a corresponding clean Lab waveform, enabling paired reference-based evaluation. However, because the mixture-generation details are unavailable, these results should not be interpreted as representative of any specific scanner sequence, noise recording, or mixing configuration.

We report the reference-free predictors and the embedding/phone measures above, together with short-time objective intelligibility (STOI)~\cite{Taal_2011}, extended STOI (ESTOI)~\cite{Jensen_2016}, and perceptual evaluation of speech quality (PESQ)~\cite{Rix_2001}, which compare each waveform with its clean Lab source. 
The probe contains 370 source utterances; LPS was evaluable for 369 utterances in both the clean-input and archived additive-noise probes, whereas the other reported metrics were evaluable for all 370. 
All condition waveforms were mono and sampled at $16$ kHz. 
Across the four processed conditions (Noisy, Denoiser, PASE, and RE-USE), 364 waveforms had the same length as their clean references. For the remaining six utterances, the waveform in every processed condition was one sample shorter than its clean reference; each enhanced waveform nevertheless had exactly the same number of samples as its corresponding noisy input. 
Before reference-based metrics were computed, each processed--reference pair was truncated to the shorter waveform, and no delay correction was applied.

We subsequently audited whole-file alignment within a $\pm0.5$-s search window. 
For Denoiser and RE-USE, we estimated the lag relative to the corresponding noisy input from the maximum absolute generalized cross-correlation with phase transform (GCC-PHAT). 
Fine-scale waveform peaks were weak or ambiguous for the vocoder-based PASE output, so for PASE we instead maximized normalized cross-correlation between centered log-root-mean-square (RMS) envelopes of the enhanced waveform and its clean source (20-ms window, 1-ms hop); 10- and 40-ms windows provided a sensitivity check.
Positive lag denotes an enhanced output delayed relative to its comparison waveform. 
This audit screens for a global buffering offset but does not detect local nonlinear timing changes or establish alignment with the rtMRI frame clock. 

\subsubsection{Acoustic modulation}
We characterize how enhancement changes two acoustic trajectories by measuring their temporal alignment between each enhanced output and its corresponding pre-enhancement input.
This is an input--output consistency analysis, not a direct test of linguistic- or paralinguistic-information preservation.

To derive a positive amplitude-change trajectory, amplitude envelopes extracted from Bark-scale critical-band-filtered signals were represented at $1000$ Hz and smoothed with zero-phase low-pass filters (10-Hz cutoff).
We then calculated the temporal derivative and set negative values to zero~\cite{Hsu_2025}.
To derive a frame-to-frame cepstral-change trajectory, we calculated the sum of squared differences in 12 mel-frequency cepstral coefficients across successive frames and smoothed the trajectory with a ninth-order (12 Hz) Butterworth low-pass filter ~\cite{Goldstein_2019}.
We then calculated phase-locking values (PLVs; values closer to 1 indicate greater alignment) between the extracted phases of each enhanced signal and its pre-enhancement input (for more details on calculating PLVs, see~\citealt{Lancia_2023}). For this analysis, the window width parameter was set to 50 ms with a step size of 5 ms, and the maximum value for the ratio between two signals' frequencies was set to 2. PLVs represent the first term in a Fourier series of the generalized phase differences, which results in one value for every window; these values across every window are then averaged to determine the PLV for each sentence.

\subsubsection{Formants}

The first three formants (F1--F3) were extracted from all audio files in the original Lab audio in TIMIT and all variations of the LSS audio using the \texttt{new-fave}\footnote{\url{https://forced-alignment-and-vowel-extraction.github.io/new-fave/}} library. Given input audio and aligned TextGrids, this library performs a series of linear predictive coding (LPC) analyses across a range of maximum-formant ceilings, yielding candidate formant trajectories sampled every 2 ms. A low-order discrete cosine transform (DCT) is fit on each candidate, with the best smoothed fit candidate being selected. We use the Lab audio as a baseline and compare it to the various forms of rtMRI audio from the LSS dataset. We subset the LSS dataset to include only the same sentences used in TIMIT. With the resulting formant tracks, we perform two analyses:
\begin{itemize}
    \item \textbf{Vowel separability}.
    As a measure of consistency in formant tracking, we use machine ABX~\cite{Schatz_2016} to test if vowel discriminability remains stable across audio conditions. Drops in ABX accuracy may indicate that enhancement causes systematic shifts in formant tracking. 
    ABX discrimination was scored on the DCT-smoothed F1--F3 tracks. Each frame's three formant values were floored at $10^{-8}$ and normalized to unit $L_1$ norm to form a 3-bin distribution, and the token-to-token distance was length-normalized dynamic time warping (DTW) whose local cost was the pairwise symmetric Kullback--Leibler (KL) divergence between frames. Within each speaker, triplets drew $A$ and $X$ from the same vowel ($X \neq A$) and $B$ from a different vowel, capped at 5 per cell and sampled without replacement, yielding a total of 210 vowel pairs and 1,050 triplets. A single shared triplet set, formed on the token intersection across conditions, was scored against every condition so the comparison is paired. A triplet scored 1 if $d(X,A) < d(X,B)$, 0 otherwise, and
    0.5 on ties (chance $= 0.5$).
   \item \textbf{Formant tracking stability}. As a simple, reference-free measure of track stability, we compute, per vowel token, the second difference of the raw (non-DCT-smoothed) F1--F3 trajectories in log-Hz, take the median of its absolute value over frames, and average across formants; we then report the distribution of this per-token value. Using the median (rather than the RMS) makes the measure robust to occasional gross tracking errors. Smooth formant motion yields near-zero curvature, so larger values indicate noisier frame-to-frame tracking.
\end{itemize}

\subsection{Downstream tasks and evaluation metrics}

\subsubsection{Automatic speech recognition}
We evaluate three ASR systems with different model and decoding architectures to test whether enhancement effects generalize across recognizers.

\begin{itemize}
    \item \textbf{Whisper Large v3}~\cite{Radford_2023}: a Transformer encoder--decoder ASR model trained with large-scale weak supervision.
    \item \textbf{Cohere Transcribe 03-2026}~\cite{Mack_2026}: a 2B-parameter, open-weight ASR model with a Conformer-based encoder and Transformer decoder.
    \item \textbf{Qwen3-ASR 1.7B}~\cite{Shi_2026}: an audio--language model that maps audio-encoder representations through a projector to an autoregressive language model.
\end{itemize}

Transcription accuracy is evaluated using word error rate (WER), the word-level edit distance between the hypothesis and reference, expressed as a percentage: $\mathrm{WER}=100(S+D+I)/N$, where $S$, $D$, and $I$ denote substitutions, deletions, and insertions, and $N$ is the number of reference words. 
We lowercase references and hypotheses, remove punctuation while preserving apostrophes, collapse whitespace, and exclude empty references. 
Counts are pooled across the evaluation set, so the reported WER is word weighted rather than an unweighted mean of utterance-level WERs. 
Lower values indicate fewer word errors.

All recordings were decoded as mono, 16-kHz audio. 
Whisper Large v3\footnote{We use repository revision \path{06f233fe06e7}.} used half-precision inference, English task conditioning, the transcription task, one-beam decoding, a 440-token limit, and timestamp-enabled long-form decoding.
Cohere Transcribe\footnote{Cohere Transcribe 03-2026 uses artifact-equivalent repository revisions \texttt{d263bc2f}\allowbreak\texttt{a85c} and \texttt{b1eacc26}\allowbreak\texttt{86a3}.} used the native Transformers implementation in bfloat16, English conditioning, one-beam decoding, and a 256-token limit.
Qwen3-ASR 1.7B used the official \texttt{qwen-asr} Transformers backend (runtime version reconstructed as 0.0.6) in half precision with English conditioning, a 512-token generation limit, and an internal batch cap of eight; no context or timestamps were requested, and its long-audio splitter was not invoked for these recordings.

\paragraph{Evaluation data.}
We run ASR on the corpora that provide verbatim transcripts. 
The evaluation sets contain 71 files from LSS (1 speaker), 911 from TIMIT (10 speakers), 1,022 from 75-Spk (75 speakers), 2,304 from EMO-MRI (10 speakers), and 366 from Child (14 speakers). 
The TIMIT set excludes three files marked as having missing sentence identifiers in the corpus manifest. 
For the 75-Spk corpus, only the stated subset is suitable for ASR. 
The recording protocol also included nonlexical vocal materials (e.g., nonsense vowel--consonant--vowel sequences and sustained sounds) and articulatory postures, which lack conventional word- or sentence-level orthographic references. 
We therefore keep only two content types---\emph{words} (lists of isolated, unrelated words) and \emph{sentences} (full reading passages)---and compute Overall results as well as separate Words and Sentences scores. 
The Child corpus follows the same words-vs-sentences split and is additionally analyzed by age group (Children vs.\ Adult Controls).

\subsubsection{Speech emotion recognition}
The EMO-MRI corpus was recorded by asking 10 professional actors to read a fixed set of sentences under four acted emotions (neutral, angry, happy, \& sad), making it suitable for a four-class speech emotion recognition (SER) probe. 
The reference for each utterance is the corpus-provided final emotion label; the corpus documentation states that this label was determined by its listener-based emotion-quality evaluation rather than taken directly from the actor's intended prompt. 
We apply the pretrained emotion2vec\_plus\_large~\cite{Ma_2024}\footnote{We use the \texttt{iic/emotion2vec\_plus\_large} model (revision \texttt{6c303ba987b8}); see \url{https://github.com/ddlBoJack/emotion2vec}.} directly, without fine-tuning. The model produces scores for nine output classes. For a closed-set four-class prediction, we retain the scores for neutral, angry, happy, and sad and select their maximum (scores for disgusted, fearful, other, surprised, and unknown are excluded). We report Macro-F1, unweighted average recall (UAR), overall accuracy, and per-emotion F1.

\subsubsection{Sex/gender classification}
We apply two pretrained models off the shelf, without fine-tuning: the gender output of the wav2vec2-large-robust age--gender model (w2v2)~\cite{Burkhardt_2023}\footnote{\url{https://huggingface.co/audeering/wav2vec2-large-robust-24-ft-age-gender}} and the sex output of the WavLM-large age--sex model from Vox-Profile (WavLM)~\cite{Feng_2025,Feng_2025_v2}\footnote{\url{https://huggingface.co/tiantiaf/wavlm-large-age-sex}}. 
We retain the terminology used by the source models but evaluate against the corpus-provided sex labels; these labels should not be interpreted as self-identified gender. 
For EMO-MRI, 75-Spk, and TIMIT, the task is binary female--male classification. 
For Child, the same model outputs are combined with the estimated age category to form four joint classes: female child, male child, female adult, and male adult. 
The Child score is therefore not directly comparable to the binary adult-corpus scores. 
LSS is excluded because it contains only one male speaker. 
We report Macro-F1 with 95\% participant-cluster bootstrap intervals. 
The TIMIT evaluation is restricted to 914 MRI-session recordings from 10 speakers; the 370 separate Lab/EMA-session recordings are excluded so that the reported condition corresponds to rtMRI audio.

\subsubsection{Age regression}
We use two models to estimate speaker age. The first is a wav2vec2-large-robust age model (hereafter w2v2)~\cite{Burkhardt_2023}\footnote{\url{https://huggingface.co/audeering/wav2vec2-large-robust-24-ft-age-gender}}. The second is the WavLM-large age model from Vox-Profile~\cite{Feng_2025,Feng_2025_v2}\footnote{The latter uses Hugging Face revision \path{a4ad8039d8e2}.}. Both are applied without fine-tuning. Our primary metric is the concordance correlation coefficient (CCC)~\cite{Lin_1989}, which captures correlation and agreement.
We also report mean absolute error (MAE, years) and mean signed error (years; positive denotes over-estimation). 
Pearson's $r$ and root-mean-square error are omitted for brevity. We evaluate 75-Spk and Child. 
Primary metrics are computed over recordings, so speakers with more recordings receive greater weight. 
As a sensitivity analysis, we average predictions within participant and recompute CCC, MAE, and mean signed error across equally weighted participant means.

\subsection{Statistical analysis}
All reported 95\% percentile-bootstrap confidence intervals use 5,000 resamples~\cite{Ferrer_2023}. 
For ASR, multi-speaker corpora resample speakers as clusters, whereas LSS resamples recordings. 
SER, sex/gender-classification, and first-language intervals use two-stage resampling: speakers or participants first, then recordings within each sampled person.

For SHEET, multi-speaker corpora used two-stage bootstrap resampling: speakers were sampled first, followed by recordings within each sampled speaker. 
Because LSS contains only one speaker, its intervals used recording-level resampling. 
This procedure was applied to both the mean SHEET score for each condition and the paired enhanced-minus-input difference. The intervals were computed using the \texttt{ConfidenceIntervals} implementation~\cite{Ferrer_2023}.
 
Quality point estimates and primary age metrics are computed over pooled recordings, so speakers with more recordings receive greater weight; speaker-balanced age intervals instead use paired two-stage resampling of participants and recordings.

\section{Experimental results and analyses}

\subsection{RQ1: Predicted quality and signal-property probes}

RQ1 regarding audio quality combines three complementary forms of evidence. Learned predictors assess apparent quality without a reference. The clean-input and archived additive-noise probes use the same Lab waveform to characterize processing-induced change and paired-reference agreement. PLV and the LSS formant analyses instead characterize input--output consistency or estimator behavior on naturally recorded rtMRI speech and are therefore interpreted as construct-specific diagnostics rather than evidence of source recovery.

\subsubsection{LSS Raw-input quality}
Table~\ref{tab:lss-raw-quality} reports the non-intrusive quality scores for the LSS Raw input and the three outputs enhanced from that input.
All three enhancers raise every score relative to Raw. The size of the gain and the model ranking nevertheless depend on the predictor: RE-USE is highest on UTMOS ($3.725$) and SHEET ($3.667$), whereas PASE is highest on deterministic single-crop UTMOSv2 ($2.833$) and VQScore ($0.744$).
These reference-free gains motivate the fidelity probes explored below rather than, by themselves, establishing preservation of the original speech signal. 
For UTMOSv2, the paired enhanced-minus-Raw differences are $+0.15\pm0.12$ for Denoiser, $+1.80\pm0.14$ for PASE, and $+1.59\pm0.15$ for RE-USE; all three pointwise intervals exclude zero.
For SHEET, all three paired intervals exclude zero.

\begin{table}[!t]
  \centering
  \caption{LSS Raw-input quality scores ($\uparrow$). Superscripts are 95\% bootstrap CI half-widths. Ranks follow Table~\ref{tab:datasets}.}
  \label{tab:lss-raw-quality}
  \setlength{\tabcolsep}{1pt}
  \begin{tabular*}{\columnwidth}{@{\extracolsep{\fill}}l cccc@{}}
    \toprule
    Metric ($\uparrow$) & Raw & Denoiser & PASE & RE-USE \\
    \midrule
    UTMOS   & 1.28\ci{<0.01} & 1.59\ci{0.02} & \underline{3.51\ci{0.06}} & \textbf{3.73\ci{0.05}} \\
    UTMOSv2\textsuperscript{a} & 1.03\ci{0.10} & 1.18\ci{0.07} & \textbf{2.83\ci{0.11}} & \underline{2.63\ci{0.12}} \\
    VQScore & 0.64\ci{<0.01} & 0.72\ci{<0.01} & \underline{0.74\ci{<0.01}} & \textbf{0.74\ci{<0.01}} \\
    SHEET\textsuperscript{b}   & 1.35\ci{0.01} & 1.70\ci{0.02} & \underline{3.11\ci{0.05}} & \textbf{3.67\ci{0.03}} \\
    \bottomrule
  \end{tabular*}
  \begin{minipage}{\columnwidth}
    \footnotesize\raggedright
    \textsuperscript{a}UTMOSv2 uses the deterministic single-crop protocol described in Sec.~\ref{ss:sq_metrics}.\par
    \textsuperscript{b}SHEET uses the deterministic long-file protocol described in Sec.~\ref{ss:sq_metrics}.
  \end{minipage}
\end{table}

\subsubsection{Clean-input probe}
To characterize processing effects on already-clean speech, we next apply each speech enhancement (SE) model directly to the Lab audio of the four TIMIT speakers (Table~\ref{tab:spksim-clean}).
Three observations stand out at the level of point estimates.
First, recognized-phone sequence agreement remains high for every model (LPS $\geq 0.95$), whereas SpkSim is lower for Denoiser ($0.720$) and PASE ($0.765$) than for RE-USE ($0.863$). 
These values indicate model-dependent speaker-embedding drift; without a calibrated impostor distribution or perceptual identity judgment, they do not establish loss of perceived speaker identity.
Second, processed clean speech can receive higher predictor scores than unprocessed Lab speech (e.g., UTMOS is $3.979$ for RE-USE and $3.849$ for Denoiser, compared with $3.655$ for Lab; SHEET is $4.121$ for RE-USE, compared with $3.749$ for Lab; RE-USE VQScore is $0.728$, compared with $0.687$ for Lab).
Learned predictors can therefore reward processing-induced changes, and a higher predicted score is not evidence that the waveform is more faithful to its source.
Third, predictor rankings are metric dependent: UTMOS and SHEET favor RE-USE, whereas the deterministic single-crop UTMOSv2 protocol favors PASE ($2.930$).

\begin{table}[!t]
  \centering
  \caption{Clean-input probe on four TIMIT speakers ($\uparrow$); SpkSim/LPS reference is unprocessed Lab (``--''). Ranks follow Table~\ref{tab:datasets}.}
  \label{tab:spksim-clean}
  \setlength{\tabcolsep}{1pt}
  \begin{tabular*}{\columnwidth}{@{\extracolsep{\fill}}l cccc@{}}
    \toprule
    Metric ($\uparrow$) & Denoiser & PASE & RE-USE & Lab \\
    \midrule
    SpkSim   & 0.72\ci{0.14} & \underline{0.77\ci{0.10}} & \textbf{0.86\ci{0.05}} & -- \\
    LPS      & \textbf{0.97\ci{0.02}} & 0.95\ci{0.02} & \underline{0.96\ci{0.02}} & -- \\ \midrule
    UTMOS    & \underline{3.85\ci{0.24}} & 3.67\ci{0.31} & \textbf{3.98\ci{0.08}} & 3.66\ci{0.42} \\
    UTMOSv2  & 2.18\ci{0.31} & \textbf{2.93\ci{0.20}} & \underline{2.65\ci{0.12}} & 2.16\ci{0.36} \\
    VQScore  & \textbf{0.73\ci{0.01}} & \underline{0.71\ci{0.01}} & \textbf{0.73\ci{0.01}} & 0.69\ci{0.01} \\
    SHEET    & \underline{3.99\ci{0.27}} & 3.97\ci{0.24} & \textbf{4.12\ci{0.16}} & 3.75\ci{0.32} \\
    \bottomrule
  \end{tabular*}
\end{table}

\subsubsection{Archived additive-noise probe}
We next evaluate the archived rtMRI-style additive mixture described above. 
Because the underlying Lab waveform is known, STOI, ESTOI, and PESQ can be computed against it, subject to the timing limitations noted above.
Table~\ref{tab:spksim-noisy} reports the unenhanced mixture (\emph{Noisy}) and three enhanced conditions. 
Relative to Noisy, Denoiser raises the ESTOI, STOI, and PESQ point estimates from $0.311$ to $0.361$, from $0.491$ to $0.539$, and from $1.433$ to $1.571$, respectively. 
The corresponding paired differences are $+0.05\pm0.06$, $+0.05\pm0.01$, and $+0.14\pm0.22$; thus, only the STOI interval excludes zero. 
PASE and RE-USE increase all three intrusive metrics relative to Noisy, with all six paired intervals excluding zero. 

For SpkSim, the paired differences are $-0.30\pm0.05$ for Denoiser, $+0.00\pm0.06$ for PASE, and $+0.09\pm0.07$ for RE-USE. 
All three models increase LPS, including Denoiser ($+0.20\pm0.10$), with paired intervals excluding zero. 
RE-USE achieved the highest point estimates across all five reference-based metrics as well as SHEET ($3.648$), whereas PASE scored slightly higher than RE-USE on UTMOS ($3.484$ vs.\ $3.459$). Because direct enhancer-to-enhancer contrast intervals were not computed, these model rankings remain descriptive.
Because the metrics have different scales and target constructs, their raw score differences are not directly comparable.

The post hoc global-alignment audit found a zero-sample GCC-PHAT lag for all 370 Denoiser and all 370 RE-USE outputs relative to their noisy inputs. 
For PASE, the log-RMS-envelope lag relative to the clean source had a median of 0 samples, an interquartile range of $-2$ to $+2$ samples and a range of $-24$ to $+12$ samples ($-1.50$ to $+0.75$ ms); estimates from 10-, 20-, and 40-ms windows differed by no more than 1 ms for every file. 
We therefore found no evidence of a systematic whole-file buffering offset in this probe. This audit does not exclude local time warping or establish frame-level audio--rtMRI synchronization.

\begin{table}[!t]
  \centering
  \caption{Additive-noise probe on four TIMIT speakers ($\uparrow$). Ranks follow Table~\ref{tab:datasets}.}
  \label{tab:spksim-noisy}
  \setlength{\tabcolsep}{1pt}
  \begin{tabular*}{\columnwidth}{@{\extracolsep{\fill}}l cccc@{}}
    \toprule
    Metric ($\uparrow$) & Noisy & Denoiser & PASE & RE-USE \\
    \midrule
    SpkSim   & \underline{0.45\ci{0.09}} & 0.15\ci{0.09} & \underline{0.45\ci{0.13}} & \textbf{0.54\ci{0.14}} \\
    LPS      & 0.11\ci{0.04} & 0.31\ci{0.13} & \underline{0.72\ci{0.15}} & \textbf{0.73\ci{0.16}} \\ \midrule
    UTMOS    & 1.32\ci{0.03} & 1.59\ci{0.26} & \textbf{3.48\ci{0.47}} & \underline{3.46\ci{0.21}} \\
    UTMOSv2  & 1.55\ci{0.10} & 1.36\ci{0.36} & \underline{2.67\ci{0.25}} & \textbf{2.73\ci{0.17}} \\
    VQScore  & 0.62\ci{0.01} & 0.71\ci{0.01} & \underline{0.73\ci{0.01}} & \textbf{0.74\ci{0.01}} \\
    SHEET    & 1.37\ci{0.06} & 1.62\ci{0.17} & \underline{3.35\ci{0.44}} & \textbf{3.65\ci{0.08}} \\ \midrule
    ESTOI    & 0.31\ci{0.09} & 0.36\ci{0.15} & \underline{0.46\ci{0.13}} & \textbf{0.54\ci{0.15}} \\
    STOI     & 0.49\ci{0.12} & 0.54\ci{0.13} & \underline{0.60\ci{0.13}} & \textbf{0.66\ci{0.13}} \\
    PESQ     & 1.43\ci{0.06} & 1.57\ci{0.24} & \underline{1.87\ci{0.33}} & \textbf{2.05\ci{0.50}} \\
    \bottomrule
  \end{tabular*}
\end{table}

\subsubsection{Acoustic modulation}

We report input--output comparisons of modulation for TIMIT-DSP, LSS-DSP, and LSS-Raw. 
For each comparison, PLVs were calculated between the input audio (Raw or DSP) and the corresponding model-enhanced output. 
Table~\ref{tab:plvs} reports median PLVs between the amplitude- and cepstral-change trajectories of each input and enhanced output.
Across all cells in Table~\ref{tab:plvs}, the median amplitude- and cepstral-change PLVs are consistently high. These values suggest that none of the three models substantially disrupts the timing of the amplitude and cepstral modulation patterns present in the input, although PLV cannot distinguish preserved speech modulation from preserved scanner-noise modulation.

All cells are descriptive input--output comparisons. The earlier analysis used independent-samples Welch tests despite repeated processing of the same recordings; we therefore omit those tests pending a paired, speaker-aware reanalysis. In particular, LSS contains one speaker and cannot support population-level claims across speakers.

\begin{table*}[!t]
  \centering
  \caption{Median input--output PLV for amplitude (\textbf{Amp}) and cepstral (\textbf{Cep.}) change under Raw/DSP inputs ($\uparrow$: more similar, not necessarily cleaner). Ranks follow Table~\ref{tab:datasets}.}
  \label{tab:plvs}
  \begin{tabular}{l cc cc cc}
    \toprule
    & \multicolumn{2}{c}{Denoiser} & \multicolumn{2}{c}{PASE} & \multicolumn{2}{c}{RE-USE} \\
    \cmidrule(lr){2-3}\cmidrule(lr){4-5}\cmidrule(lr){6-7}
    Audio & Amp & Cep. & Amp & Cep. & Amp & Cep. \\
    \midrule
    TIMIT-DSP & \textbf{0.99} & \textbf{0.93} & 0.98 & 0.90 & \underline{0.98} & \underline{0.91} \\
    LSS-DSP & \textbf{0.96} & \textbf{0.88} & 0.96 & 0.87 & \underline{0.96} & \underline{0.88} \\
    LSS-Raw & \textbf{0.94} & \textbf{0.94} & \underline{0.92} & \underline{0.92} & 0.92 & 0.91 \\
    \bottomrule
  \end{tabular}
\end{table*}

\subsubsection{Formants}
\begin{figure*}[!t]
    \centering
    \includegraphics[width=\textwidth,height=0.72\textheight,keepaspectratio]{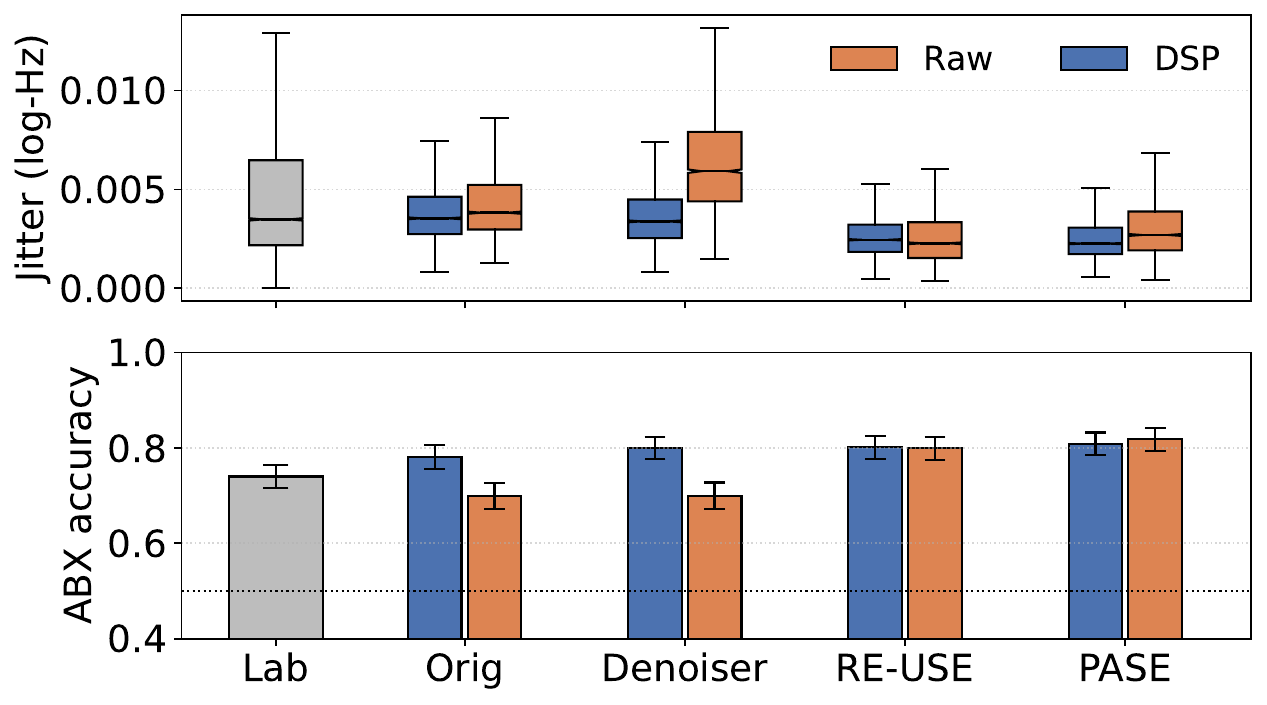}
    \caption{Formant-trajectory ABX scores (bottom) and the formant-track stability (jitter) measure (top) for unenhanced and model-enhanced Raw and DSP inputs. The unpaired USC-TIMIT Lab acoustic comparator is labeled \emph{Lab} in the panel, and unenhanced input is labeled \emph{Orig.} The dotted line at 0.5 marks ABX chance. ABX uses DCT-smoothed F1--F3 tracks; the upper panel uses raw F1--F3 tracks. Lower values indicate smoother estimated trajectories.} 
    \label{fig:abx}
\end{figure*}

Figure~\ref{fig:abx} reports formant-trajectory ABX scores (bottom) and formant-tracking stability, termed jitter (top), for unenhanced and enhanced Raw and DSP inputs. All reported ABX means are at least 0.70, well above chance. All rtMRI conditions except unenhanced Raw and Denoiser-enhanced Raw are near or above the unpaired Lab comparator (0.70 for both versus 0.74 for Lab). Paired Wilcoxon tests indicate higher ABX scores for DSP than Raw in the unenhanced and Denoiser conditions ($p < 0.01$); the corresponding Raw--DSP difference is negligible for RE-USE and PASE. For the Raw input, PASE and RE-USE have significantly higher ABX scores than unenhanced Raw ($p < 0.001$), whereas no enhancer significantly exceeds unenhanced DSP for the DSP input. 
Thus, the ABX effect is input- and model-dependent. PASE and RE-USE improve vowel discriminability relative to unenhanced Raw, whereas no enhancer provides a detectable additional gain over unenhanced DSP.
\par
For the upper-panel formant-track stability measure (jitter), lower values denote smoother estimated tracks. Paired Wilcoxon tests indicate lower values for DSP than Raw in every condition except RE-USE ($p < 0.001$). Denoiser applied to Raw has the largest value among the displayed conditions. Relative to the corresponding unenhanced input, PASE and RE-USE have lower values on the Raw input, and all three enhancers have lower values on the DSP input ($p < 0.001$). The results across ABX and jitter show some disconnect. While little improvement is seen in ABX when enhancing beyond DSP processing, enhancement can improve jitter beyond DSP. Similarly, there were no differences between DSP and Raw inputs for PASE and RE-USE in ABX, but for jitter, differences do emerge -- RE-USE shows more stability given the Raw input, while PASE shows the opposite. This highlights the targeted distinction between formant-tracking stability and consistency in formant placement. A given formant trajectory can be placed relatively consistently across conditions while also exhibiting varying degrees of jitter at local timescales.

Formant-estimation agreement in the archived additive-noise probe is reported in Table~\ref{tab:formant-error}. 
We use formant deviation/gross-error rate (FDR) and formant estimation error (FEE) as metrics of formant preservation~\cite{Alku_2023,Kadiri_2025}. FDR is the percentage of the evaluated duration for which the estimated formant deviates from the clean-audio estimate by more than 30\%; lower is better. FEE is the mean absolute frequency deviation in Hz; lower is better.
Relative to Noisy, Denoiser has lower F1 and F2 FDR point estimates but a higher F3 FDR estimate, while its FEE estimates are lower for all three formants. PASE and RE-USE have lower FDR and FEE point estimates for F1--F3. These comparisons indicate greater agreement with automatic clean-waveform estimates, not recovery of manually verified ground-truth formants.

\begin{table}[!b]
  \centering
  \caption{Automatic-formant agreement in the additive-noise probe: FDR (30\% gross-error threshold) and FEE (Hz), both $\downarrow$. Ranks follow Table~\ref{tab:datasets}.}
  \label{tab:formant-error}
  \footnotesize
  \setlength{\tabcolsep}{0pt}
  \begin{tabular*}{\columnwidth}{@{\extracolsep{\fill}}llcccc@{}}
    \toprule
    Metric ($\downarrow$) & F & Noisy & Denoiser & PASE & RE-USE \\
    \midrule
    \multirow{3}{*}{FDR (\%)} & F1 & 79.01\ci{1.58} & \underline{18.55~\ci{0.59}} & \textbf{14.67\ci{0.61}} & \textbf{14.67\ci{0.59}} \\
     & F2 & 33.75\ci{1.93} & 19.21~\ci{0.71} & \underline{11.87\ci{0.72}} & \textbf{11.50\ci{0.85}} \\
     & F3 & 12.47\ci{1.16} & 16.19~\ci{0.49} & \textbf{7.87\ci{0.65}} & \underline{7.95\ci{0.65}} \\
    \midrule
    \multirow{3}{*}{FEE (Hz)} & F1 & 939.05\ci{23.90} & 86.26~\ci{4.47} & \underline{74.23\ci{3.34}} & \textbf{72.64\ci{6.44}} \\
     & F2 & 1391.65\ci{28.51} & 196.01~\ci{9.82} & \underline{149.60\ci{7.28}} & \textbf{140.22\ci{14.25}} \\
     & F3 & 1648.48\ci{25.64} & 266.84~\ci{13.60} & \underline{200.29\ci{9.48}} & \textbf{196.24\ci{19.78}} \\
    \bottomrule
  \end{tabular*}
\end{table}

\subsection{RQ2: Automatic speech recognition}
WERs for three ASR systems on the unenhanced inputs and their three model-enhanced variants are reported in Tables~\ref{tab:asr-1} and~\ref{tab:asr-2}.
For LSS, both the Raw-input and DSP-input branches are shown; the remaining corpora use DSP as the input. 

\begin{table*}[!t]
  \centering
  \caption{Word-weighted ASR WER (\%; $\downarrow$) for unenhanced and enhanced LSS (Raw/DSP) and EMO-MRI (DSP). Ranks follow Table~\ref{tab:datasets}.}
  \label{tab:asr-1}
  \begin{tabular*}{\textwidth}{@{\extracolsep{\fill}}llcccc@{}}
    \toprule
    Corpus & ASR & Unenh. & Denoiser & PASE & RE-USE \\
    \midrule
    \multirow{3}{*}{LSS (Raw)} & Qwen3 & \underline{6.29\,\ci{0.99}} & 10.02\,\ci{1.36} & 6.60\,\ci{0.93} & \textbf{4.90\,\ci{0.83}} \\
     & Whisper & 11.35\,\ci{6.78} & 9.62\,\ci{1.28} & \underline{8.48\,\ci{1.28}} & \textbf{6.05\,\ci{1.13}} \\
     & Cohere & \textbf{5.57\,\ci{1.00}} & 7.77\,\ci{1.10} & 7.80\,\ci{1.04} & \underline{5.60\,\ci{0.93}} \\
    \midrule
    \multirow{3}{*}{LSS (DSP)} & Qwen3 & \textbf{4.33\,\ci{0.76}} & 5.03\,\ci{0.91} & 4.84\,\ci{0.79} & \underline{4.58\,\ci{0.83}} \\
     & Whisper & \underline{6.09\,\ci{1.09}} & 6.20\,\ci{1.11} & \textbf{6.01\,\ci{1.16}} & 6.23\,\ci{1.68} \\
     & Cohere & 6.05\,\ci{1.03} & \underline{5.91\,\ci{1.05}} & 6.26\,\ci{1.12} & \textbf{5.46\,\ci{0.97}} \\
    \midrule
    \multirow{3}{*}{EMO-MRI} & Qwen3 & \textbf{6.08\,\ci{2.36}} & 9.61\,\ci{3.69} & 7.31\,\ci{2.71} & \underline{6.71\,\ci{2.50}} \\
     & Whisper & 30.70\,\ci{7.91} & 34.29\,\ci{9.96} & \textbf{25.94\,\ci{9.92}} & \underline{28.65\,\ci{6.37}} \\
     & Cohere & 21.61\,\ci{6.30} & 29.90\,\ci{10.12} & \textbf{19.60\,\ci{9.59}} & \underline{20.77\,\ci{9.95}} \\
    \bottomrule
  \end{tabular*}
\end{table*}

\begin{table*}[!t]
  \centering
  \caption{ASR WER for TIMIT, 75-Spk, and Child DSP inputs (word-weighted \%; $\downarrow$). TIMIT excludes Lab; Child pools cohorts (Table~\ref{tab:child-asr-age}). Ranks follow Table~\ref{tab:datasets}.}
  \label{tab:asr-2}
  \begin{tabular*}{\textwidth}{@{\extracolsep{\fill}}llcccc@{}}
    \toprule
    Corpus & ASR & Unenh. & Denoiser & PASE & RE-USE \\
    \midrule
    \multirow{3}{*}{TIMIT} & Qwen3 & \underline{6.86\,\ci{3.64}} & 11.82\,\ci{8.08} & 7.43\,\ci{3.73} & \textbf{6.63\,\ci{3.52}} \\
     & Whisper & \underline{7.32\,\ci{3.64}} & 15.33\,\ci{11.41} & 8.05\,\ci{3.77} & \textbf{7.06\,\ci{3.53}} \\
     & Cohere & \underline{6.58\,\ci{3.56}} & 14.25\,\ci{11.72} & 7.41\,\ci{3.76} & \textbf{6.44\,\ci{3.54}} \\
    \midrule
    \multirow{3}{*}{75-Spk} & Qwen3 & \underline{10.87\,\ci{2.67}} & 12.99\,\ci{3.11} & 10.97\,\ci{2.47} & \textbf{10.70\,\ci{2.38}} \\
     & Whisper & 12.47\,\ci{2.78} & 15.90\,\ci{3.59} & \textbf{12.15\,\ci{2.90}} & \underline{12.28\,\ci{3.13}} \\
     & Cohere & \textbf{9.93\,\ci{2.33}} & 12.57\,\ci{3.36} & 10.29\,\ci{2.39} & \underline{10.04\,\ci{2.27}} \\
    \midrule
    \multirow{3}{*}{\shortstack{Child\\Overall}} & Qwen3 & \underline{27.56\,\ci{8.93}} & 31.93\,\ci{11.30} & 28.96\,\ci{9.41} & \textbf{25.89\,\ci{8.91}} \\
     & Whisper & 46.46\,\ci{12.03} & 38.58\,\ci{11.19} & \textbf{32.26\,\ci{8.82}} & \underline{33.62\,\ci{10.45}} \\
     & Cohere & \underline{30.41\,\ci{11.95}} & 34.61\,\ci{12.38} & \underline{30.41\,\ci{9.83}} & \textbf{29.13\,\ci{11.50}} \\
    \bottomrule
  \end{tabular*}
\end{table*}

To summarize the multi-corpus pattern relative to DSP, we define $\Delta\mathrm{WER}$ as enhanced WER minus DSP WER; negative values therefore indicate a lower point estimate. 
Across the five DSP-input corpora and three ASR systems (15 comparisons per enhancement model), RE-USE lowers WER in 11 comparisons (unweighted mean/median $\Delta\mathrm{WER}=-1.28/-0.23$ percentage points). 
PASE lowers WER in five comparisons, raises it in nine, and produces one tie at the reported precision (mean/median $=-1.03/+0.21$ points). 
The difference between PASE's mean and median reflects a small number of relatively large decreases rather than a consistent effect. Denoiser raises WER in 13 comparisons and lowers it in two (mean/median $=+3.04/+3.54$ points). 
In the paired analysis of these DSP branches, the pointwise interval excludes zero for nine of 15 Denoiser contrasts (all toward higher WER), six PASE contrasts (five higher and Child/Whisper lower), and three RE-USE contrasts (all lower).

On the separate LSS Raw branch, Denoiser raises Qwen3 WER from $6.29\%$ to $10.02\%$ (paired $\Delta=+3.73\pm1.14$) and raises Cohere WER from $5.57\%$ to $7.77\%$ ($+2.21\pm0.75$); it lowers the Whisper point estimate from $11.35\%$ to $9.62\%$, but the paired interval includes zero ($-1.73\pm6.86$). 
PASE has a lower WER point estimate only for Whisper. 
RE-USE lowers Qwen3 from $6.29\%$ to $4.90\%$ (paired $\Delta=-1.39\pm0.88$) and lowers Whisper from $11.35\%$ to $6.05\%$ ($-5.31\pm6.60$), while Cohere is essentially unchanged ($5.57\%$ to $5.60\%$; $+0.03\pm0.47$). 
Enhancement effects therefore depend on corpus, recognizer, model, and input condition. Higher non-intrusive predicted-quality scores do not reliably imply lower ASR error on rtMRI speech.

\subsection{RQ3: Paralinguistic analysis}
\subsubsection{Speech emotion recognition}

For SER (Table~\ref{tab:ser}), Denoiser, PASE, and RE-USE have higher aggregate point estimates than DSP, with RE-USE highest on Macro-F1 ($48.50\rightarrow56.83$), UAR, and accuracy.

\begin{table*}[!t]
  \centering
  \caption{Four-class SER on EMO-MRI (2,304 recordings, 10 speakers; emotion2vec+ large; \%; $\uparrow$). UAR: unweighted average recall. Ranks follow Table~\ref{tab:datasets}.}
  \label{tab:ser}
  \begin{tabular}{lcccc}
    \toprule
    Metric ($\uparrow$) & DSP & Denoiser & PASE & RE-USE \\
    \midrule
    Macro-F1 & 48.50\,\ci{6.40} & 52.54\,\ci{6.58} & \underline{53.32\,\ci{7.95}} & \textbf{56.83\,\ci{8.90}} \\
    UAR & 51.21\,\ci{5.86} & 54.56\,\ci{5.65} & \underline{54.89\,\ci{7.19}} & \textbf{58.37\,\ci{7.93}} \\
    Accuracy & 51.82\,\ci{6.05} & 55.12\,\ci{5.84} & \underline{55.43\,\ci{7.16}} & \textbf{59.07\,\ci{7.89}} \\
    \midrule
    F1-score (neutral) & 62.51\,\ci{9.14} & \underline{64.66\,\ci{9.43}} & 63.60\,\ci{9.55} & \textbf{66.98\,\ci{9.20}} \\
    F1-score (angry) & 46.91\,\ci{10.98} & \textbf{54.32\,\ci{9.68}} & 51.32\,\ci{9.56} & \underline{54.10\,\ci{10.32}} \\
    F1-score (happy) & 55.16\,\ci{6.75} & 56.37\,\ci{6.29} & \underline{57.03\,\ci{7.75}} & \textbf{61.31\,\ci{8.64}} \\
    F1-score (sad) & 29.42\,\ci{11.79} & 34.82\,\ci{15.65} & \underline{41.31\,\ci{15.67}} & \textbf{44.94\,\ci{16.84}} \\
    \bottomrule
  \end{tabular}
\end{table*}

For Denoiser, the paired enhanced-minus-DSP differences are $+4.04\pm3.38$ for Macro-F1, $+3.35\pm2.94$ for UAR, and $+3.30\pm3.04$ for accuracy; all three pointwise intervals exclude zero. 
For PASE, the Macro-F1 difference is $+4.81\pm4.89$; its original asymmetric percentile interval excludes zero even though the symmetric display approximation extends slightly below zero. The UAR ($+3.68\pm4.10$) and accuracy ($+3.60\pm4.00$) intervals include zero. 
For RE-USE, the differences are $+8.33\pm5.26$ for Macro-F1, $+7.16\pm4.63$ for UAR, and $+7.25\pm4.51$ for accuracy; all three intervals exclude zero. 
The largest per-class point-estimate change is for \emph{sad} ($29.42\rightarrow44.94$)~\footnote{In the model's unrestricted nine-class output, the top-scoring label falls outside the four corpus categories for $24.44\%$ of DSP recordings, $23.48\%$ after Denoiser, $20.23\%$ after PASE, and $14.06\%$ after RE-USE. 
Thus, this closed-set analysis characterizes discrimination among the four corpus categories and should not be interpreted as nine-class recognition performance.}.

\subsubsection{Sex classification}
\begin{figure*}[!t]
  \centering
  \includegraphics[width=\reprintcolumnwidth,height=0.55\textheight,keepaspectratio]{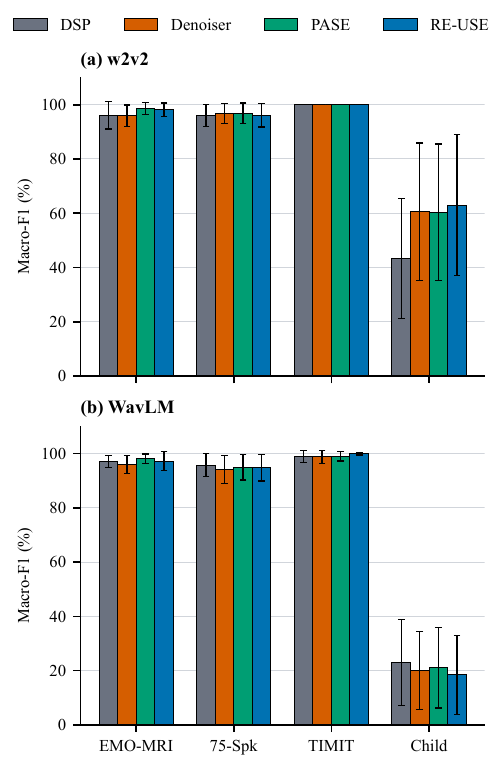}
  \caption{Macro-F1 (\%, higher is better) for the sex/gender-related classification probes. EMO-MRI, 75-Spk, and TIMIT use binary corpus-provided female--male labels; Child uses four joint sex--age-group classes and is therefore not directly comparable in task difficulty. The TIMIT panel is restricted to 914 MRI-session recordings from 10 speakers; the 370 separate Lab/EMA-session recordings are excluded. Error bars show the symmetric 95\% bootstrap-CI approximation $\pm(\mathrm{upper}-\mathrm{lower})/2$ from 5,000 two-stage participant/recording percentile-bootstrap resamples. LSS is excluded because it has one male speaker.}
  \label{fig:gender}
\end{figure*}

Figure~\ref{fig:gender} shows high binary-classification scores on the adult corpora, but the pattern should be interpreted as a probe of two pretrained representation spaces rather than as a demographic-classification recommendation. 
On the TIMIT MRI-only subset (914 recordings from 10 speakers), w2v2 achieved $100.00\pm0.00\%$ Macro-F1 under all four conditions. WavLM yielded $98.91\pm2.24\%$ on DSP, $98.80\pm2.44\%$ with Denoiser, $99.02\pm1.75\%$ with PASE, and $99.89\pm0.35\%$ with RE-USE. 
On Child's more difficult four-class joint sex--age-group task, w2v2 increases from $43.33\%$ on DSP to $60.54\%$ with Denoiser, $60.33\%$ with PASE, and $63.01\%$ with RE-USE, whereas WavLM ranges from $18.43\%$ to $23.00\%$ across conditions. 
The wide Child intervals reflect the small sample of 14 participants. Because the Adult and Child panels use different target spaces (binary sex versus joint sex \& age group), this contrast does not establish that child speech acoustics cause the lower scores. It shows only that the evaluated off-the-shelf models are unreliable for the joint labels required by this mixed Child--Adult corpus, particularly for WavLM.

\subsection{RQ4: Performance across age and first-language groups}
Enhancement effects vary across the age and first-language groups examined here. 
We probe age (Children vs.\ Adult Controls in Child corpus) and reported first-language in 75-Spk. 
For the latter analysis, first language (L1) is operationalized from the corpus metadata. The exact metadata value \texttt{L1=English} defines the L1-English group; all other L1 entries define the other-L1 group, yielding 48 and 27 speakers, respectively.
(This operational split differs from the source corpus's reported 49 native and 26 non-native speakers because the two classifications use different metadata rules.)  

\subsubsection{Age}
The age-stratified ASR results in Table~\ref{tab:child-asr-age} show a large observed Child--Adult-Control difference: for example, Qwen3 on DSP yields $33.32\%$ WER for Children and $8.26\%$ for Adult-Control. 
Subtracting the Adult-Control point estimates from the Child point estimates, the Qwen3 gap is $25.06$ WER points on DSP, $29.06$ with Denoiser, and $23.21$ with RE-USE; for Cohere, it is $22.66$ on DSP and $16.43$ with PASE. 
These gap changes are descriptive because contrast intervals were not computed. 
They also cannot be attributed to age alone: there are only three individual adult controls, and age group is confounded with other corpus and participant differences.

\begin{table*}[!t]
  \centering
  \caption{Child ASR WER by age cohort (word-weighted \%; $\downarrow$); pooled results are in Table~\ref{tab:asr-2}. Ranks follow Table~\ref{tab:datasets}.}
  \label{tab:child-asr-age}
  \begin{tabular}{c|c|cccc}
    \toprule
    Age Group & ASR & DSP & Denoiser & PASE & RE-USE \\
    \midrule
    \multirow{3}{*}{Children}
      & Qwen3   & \underline{33.32\,\ci{9.03}} & 38.61\,\ci{12.06} & 34.80\,\ci{9.73} & \textbf{31.23\,\ci{9.34}} \\
      & Whisper & 49.77\,\ci{10.38} & 44.17\,\ci{12.35} & \textbf{37.22\,\ci{9.21}} & \underline{38.52\,\ci{11.34}} \\
      & Cohere  & 35.62\,\ci{13.66} & 39.69\,\ci{14.25} & \textbf{34.18\,\ci{11.04}} & \underline{34.28\,\ci{13.13}} \\
    \midrule
    \multirow{3}{*}{Adult}
      & Qwen3   & \underline{8.26\,\ci{3.11}} & 9.55\,\ci{2.88} & 9.40\,\ci{2.83} & \textbf{8.02\,\ci{2.74}} \\
      & Whisper & 35.38\,\ci{35.28} & 19.84\,\ci{10.91} & \textbf{15.66\,\ci{8.12}} & \underline{17.20\,\ci{13.83}} \\
      & Cohere  & \underline{12.96\,\ci{7.52}} & 17.57\,\ci{13.65} & 17.75\,\ci{13.18} & \textbf{11.86\,\ci{7.24}} \\
    \bottomrule
  \end{tabular}
\end{table*}

Age regression provides a complementary probe of retained age-related information, but the Child results are reported only for the combined cohort and cannot establish a child--adult regression gap. 
Evaluable age outputs were available for 366 Child recordings from all 14 participants; the analysis subset excludes 84 corpus files whose elicitation items are absent from the age-analysis label manifest. 
On 75-Spk, age predictions were evaluable for 994 recordings from 73 speakers; 28 recordings from two speakers were excluded because their age metadata were unavailable. Because each speaker contributed multiple recordings, confidence intervals used two-stage bootstrap resampling of speakers and then recordings within each sampled speaker.

\begin{figure}[!t]
  \centering
  \includegraphics[width=\reprintcolumnwidth,height=0.55\textheight,keepaspectratio]{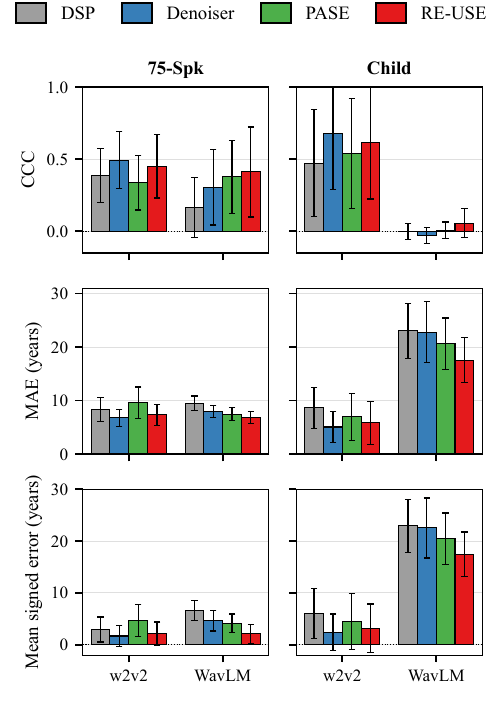}
  \caption{Results on the \emph{age regression} task across two corpora (75-Spk: 994 recordings, 73 speakers; Child: 366 recordings, 14 participants). Rows report CCC ($\uparrow$), MAE ($\downarrow$, years), and mean signed error ($\rightarrow 0$, years; positive denotes over-estimation). Point estimates aggregate recordings rather than speakers, so speakers with more recordings receive greater weight. The Child bars aggregate 11 children and 3 adult controls. On this mixed cohort, WavLM has near-zero CCC and MAE of $17.55$--$23.01$ years; nearly all absolute-error magnitude reflects over-estimation rather than under-estimation.}
  \label{fig:age}
\end{figure}

Across the two corpora (Fig.~\ref{fig:age}), effects vary by estimator and condition; for example, w2v2 CCC on 75-Spk increases from $0.386$ to $0.493$ with Denoiser. 

On the mixed-cohort Child corpus, WavLM yields near-zero CCC ($-0.030$ to $0.057$) and MAE of $17.55$--$23.01$ years. 
Its mean signed error ($17.45$--$22.96$ years) is close to its MAE, indicating that nearly all absolute-error magnitude reflects over-estimation rather than under-estimation; this does not imply an absence of dispersion across recordings or speakers. 
The estimators differ in task-specific training coverage: the WavLM estimator uses Common Voice, TIMIT, and age-enriched VoxCeleb, whereas the w2v2 estimator also uses aGender, which contains child speech~\cite{Burkhardt_2023,Feng_2025}. 
The youngest group documented for the WavLM training distribution is the teens, so the children here ($6.75$--$8.67$ years) fall below its age coverage. 
This mismatch may have contributed to the contrast, but it does not establish causality because the models also differ in architecture, upstream pretraining, and adaptation. 
Moreover, the pooled Child CCC can be driven by separation of the Child and Adult cohorts rather than sensitivity to within-child age variation.

Equal weighting of participants preserves the main pattern. In 75-Spk, the speaker-balanced CCC/MAE for w2v2 is $0.491/7.28$ years (DSP), $0.613/5.58$ (Denoiser), $0.461/8.24$ (PASE) and $0.558/6.46$ (RE-USE); the corresponding WavLM values are $0.191/9.00$, $0.380/7.22$, $0.442/7.00$, and $0.491/6.27$. 
On Child, speaker-balanced WavLM CCC remains near zero ($-0.033$ to $0.062$), while MAE and positive mean signed error remain $17.46$--$22.94$ years. 
The Child intervals are wide because only 14 participants are available, and equal participant weighting does not remove the confounding introduced by combining 11 children with 3 adult controls.

\subsubsection{First-language group}
The effect of enhancement on the WER gap between the other-L1 and L1-English groups was both model- and ASR-dependent (Table~\ref{tab:accent}).
Relative to DSP, Denoiser widened the point-estimate gap from $+4.5$ to $+6.6$ points for Qwen3, from $+4.4$ to $+7.7$ for Whisper, and from $+2.7$ to $+6.9$ for Cohere.
RE-USE narrowed the gap from $+4.5$ to $+2.7$ points for Qwen3 and from $+2.7$ to $+2.5$ for Cohere, but widened it from $+4.4$ to $+5.1$ for Whisper.
Because the table reports marginal intervals for each group rather than intervals for the between-group gaps or their changes, these patterns should be interpreted descriptively.

\begin{table}[!t]
  \centering
  \caption{75-Spk first-language-group ASR WER (word-weighted \%; $\downarrow$), pooling words/sentences; groups have 48/27 speakers. $\Delta=$ other-L1 minus L1-English WER; smaller $|\Delta|$ means a smaller observed gap (no contrast intervals). Conditions are ranked within ASR/group; ranks follow Table~\ref{tab:datasets}.}
  \label{tab:accent}
  \footnotesize
  \setlength{\tabcolsep}{1pt}
  \begin{tabular*}{\columnwidth}{@{\extracolsep{\fill}}llccc@{}}
    \toprule
    Cond. & Group & Qwen3 & Whisper & Cohere \\
    \midrule
    \multirow{3}{*}{DSP} & L1-English & \textbf{9.24\,\ci{3.20}} & 10.87\,\ci{3.83} & \textbf{8.96\,\ci{3.13}} \\
     & Other-L1 & 13.73\,\ci{5.12} & \textbf{15.30\,\ci{5.64}} & \textbf{11.64\,\ci{4.28}} \\
     & $\Delta$ & $+4.50$ & $\mathbf{+4.40}$ & $\underline{+2.70}$ \\
    \midrule
    \multirow{3}{*}{Denoiser} & L1-English & 10.61\,\ci{3.37} & 13.13\,\ci{4.23} & 10.06\,\ci{3.17} \\
     & Other-L1 & 17.18\,\ci{6.21} & 20.79\,\ci{8.04} & 16.98\,\ci{7.88} \\
     & $\Delta$ & $+6.60$ & $+7.70$ & $+6.90$ \\
    \midrule
    \multirow{3}{*}{PASE} & L1-English & \underline{9.56\,\ci{3.22}} & \textbf{9.78\,\ci{3.22}} & \underline{9.10\,\ci{3.16}} \\
     & Other-L1 & \underline{13.46\,\ci{4.58}} & 16.33\,\ci{6.47} & 12.37\,\ci{4.39} \\
     & $\Delta$ & $\underline{+3.90}$ & $+6.50$ & $+3.30$ \\
    \midrule
    \multirow{3}{*}{RE-USE} & L1-English & 9.74\,\ci{3.22} & \underline{10.42\,\ci{3.55}} & 9.12\,\ci{3.16} \\
     & Other-L1 & \textbf{12.39\,\ci{4.19}} & \underline{15.55\,\ci{7.41}} & \underline{11.65\,\ci{4.16}} \\
     & $\Delta$ & $\mathbf{+2.70}$ & $\underline{+5.10}$ & $\mathbf{+2.50}$ \\
    \bottomrule
  \end{tabular*}
\end{table}

\section{Discussion and limitations}

The central finding is that speech enhancement acts as a measurement-transforming front end rather than a uniformly beneficial cleanup step. On LSS Raw audio, all three systems increased the four learned quality scores, but their ASR effects differed across enhancers and recognizers. Across the five DSP-input corpora, Denoiser raised the WER point estimate in 13 of 15 corpus--recognizer comparisons. In the archived additive-noise probe, paired intervals supported increases from Noisy for PASE and RE-USE in recognized-phone agreement and the intrusive intelligibility and quality measures. Thus, a higher non-intrusive predicted-quality score is insufficient evidence that a waveform is better suited to a specific speech-science or machine-processing task.

The complementary probes clarify why no single metric can define overall quality. LPS measures recognized-phone agreement, SpkSim similarity in one speaker-embedding space, and ESTOI, STOI, and PESQ distinct reference-based constructs; none directly establishes listener-perceived phonetic or speaker-identity preservation. PLV measures input--output modulation consistency and may reflect speech modulation, scanner-noise modulation, or both. Formant-track curvature, ABX, and FDR/FEE respectively characterize temporal smoothness, vowel-category separability in automatically estimated F1--F3 trajectories, and agreement with automatic clean-waveform estimates; they do not establish behavioral perception or manually verified ground truth. Architecture and training data could help explain model differences, but MRI noise is out of domain for all three checkpoints and this output-level study cannot identify causal mechanisms~\cite{Defossez_2020,Rong_2026,Fu_2026}.

Downstream and speaker-group findings were also task specific. Enhancement increased several aggregate SER point estimates, while adult-corpus sex-related probes remained near ceiling. The Child joint sex--age task used a different target space and is not a controlled adult--child comparison. Age-estimation patterns varied by estimator and corpus, and the observed Child--Adult-Control and other-L1--L1-English WER gaps were descriptive. These analyses are limited by small or imbalanced participant samples, confounding among age, corpus, speech material, and participant characteristics, metadata-based first-language grouping, and unavailable gap-change intervals. They identify conditions requiring task-specific validation but do not establish causal demographic effects of enhancement.

Several additional limitations constrain generalization. 
Raw audio was available only for single-speaker LSS, whose post-DSP processing provenance could not be recovered, and the archived additive-noise probe included only four TIMIT speakers. 
Also, the alignment audit found no systematic whole-file buffering offset, and enhanced outputs retained the sample count of their noisy inputs, but it did not test local timing or frame-level audio--rtMRI/EMA correspondence. 
Original and DSP waveforms and timing information should therefore be retained, with enhanced derivatives validated for the intended acoustic, articulatory, or downstream endpoint.

\section{Conclusion and future work}
Across five rtMRI corpora with synchronized speech audio, the effects of off-the-shelf speech enhancement depend on the input representation, corpus, evaluator, and downstream task. Increases from learned reference-free quality predictors are insufficient to infer improvements in ASR or reference-based fidelity measures. The acoustic-modulation and formant probes likewise address distinct constructs: input--output modulation similarity, formant track stability, and automatic vowel separability do not by themselves establish clean-speech fidelity or listener perception. Among the evaluated checkpoints, RE-USE produces the most favorable descriptive ASR pattern, but no system is uniformly best. Enhanced rtMRI speech audio should therefore be treated as a transformed measurement channel rather than a universally improved substitute for the original signal.
For rtMRI research, DSP and learned enhancement should be treated as complementary rather than interchangeable. The original and DSP-processed waveforms should remain available as reference records, particularly for analyses requiring acoustic--articulatory synchronization or fine phonetic measurement, because this study did not establish a consistent acoustic-phonetic advantage of learned enhancement over DSP nor verify frame-level synchronization. 
A learned enhancer may be added as a task-specific derivative when matched validation demonstrates a benefit for the intended downstream analysis or application. Future work should add listener evaluation, estimate direct model-to-model contrasts, test frame-level synchronization, and evaluate in-domain adaptation.

\begin{acknowledgments}
This work was supported in part by the US National Science Foundation (IIS-2311676) and the National Science and Technology Council (NSTC), Taiwan (114-2917-I-564-030 to Huang-Cheng Chou).

AI-assisted tools were used solely for language and formatting review. All scientific content, analyses, citations, and conclusions were reviewed and verified by the authors.
\end{acknowledgments}

\section*{Author Declarations}

\subsection*{Conflict of Interest}
The authors have no conflicts to disclose.

\subsection*{Ethics Approval}
The collection and use of the Child corpus were approved by the Institutional Review Board of the University of Southern California (Protocol No.~HS-21-00712). Adult control participants provided written informed consent. For minor participants, written permission was obtained from a parent or legal guardian, and assent was obtained from each child participant.

\subsection*{Author Contributions}
\textbf{Huang-Cheng Chou}: Conceptualization (lead); Data Curation; Formal Analysis (lead); Funding Acquisition; Investigation (lead); Methodology (lead); Project Administration; Software; Visualization; Writing -- Original Draft (lead); Writing -- Review \& Editing.

\textbf{Sean Foley}: Data Curation; Formal Analysis; Investigation; Methodology; Visualization; Software; Resources; Writing -- Review \& Editing.

\textbf{Haley Hsu}: Formal Analysis; Investigation; Methodology; Software; Visualization; Writing -- Review \& Editing.

\textbf{Kevin Huang}: Formal Analysis (synthetic MRI speech); Investigation; Software; Visualization; Writing -- Review \& Editing.

\textbf{Szu-Jui Chen}: Formal Analysis (ASR); Investigation; Software; Writing -- Review \& Editing.

\textbf{Rong Chao}: Investigation (speech enhancement); Methodology; Software; Writing -- Review \& Editing.

\textbf{Louis Goldstein}: Conceptualization; Resources; Supervision; Writing -- Review \& Editing.

\textbf{Khalil Iskarous}: Conceptualization; Resources; Supervision; Writing -- Review \& Editing.

\textbf{Dani Byrd}: Conceptualization; Resources; Supervision; Writing -- Review \& Editing.

\textbf{Yu Tsao}: Conceptualization; Supervision; Writing -- Review \& Editing.

\textbf{Sudarsana Reddy Kadiri}: Supervision; Writing -- Review \& Editing.

\textbf{John H.~L.~Hansen}: Writing -- Review \& Editing.

\textbf{Shrikanth Narayanan}: Conceptualization; Funding Acquisition; Project Administration; Resources; Supervision; Writing -- Review \& Editing.
All authors reviewed and approved the final manuscript.

\section*{Data Availability}
The public corpora analyzed in this study are available as follows: LSS at \url{https://sail.usc.edu/span/single_spk/}; USC-TIMIT at \url{https://zenodo.org/records/19422914}; USC 75-Speaker at \url{https://doi.org/10.6084/m9.figshare.13725546.v1}; and USC-EMO-MRI at \url{https://doi.org/10.5281/zenodo.19325044}.

The Child corpus, archived additive-noise pairs, analysis code, sample manifests, derived scores, ASR hypotheses, and table and figure source data are available from the corresponding author, Shrikanth Narayanan (\href{mailto:shri@usc.edu}{shri@usc.edu}), upon reasonable request and subject to applicable ethical and institutional requirements.

The public interactive demonstration at \url{https://rmridemo.huangchengchou.com} provides illustrative audio examples and is not an archival data repository.

\bibliography{references}

\end{document}